# Synthesis and Stoichiometry of MgB$_2$

D.G. Hinks*, J.D. Jorgensen, Hong Zheng, and S. Short

Materials Science Division, Argonne National Laboratory, Argonne IL 60439

ABSTRACT

The system Mg$_x$B$_2$ has been studied to investigate possible nonstoichiometry in MgB$_2$. When synthesized at 850°C, MgB$_2$ is a line compound with a possible Mg vacancy content of about 1%. Small changes in lattice constants as a function of starting composition result from grain interaction stresses, whose character is different in the Mg-rich, near-stoichiometric, and Mg-deficient regimes. A small linear decrease of the superconducting transition temperature, $T_c$, in the Mg-rich regime results from accidental impurity doping.

PACS: 74.72.Yg; 74.62.Bf

Keywords: MgB2; stoichiometry; synthesis

*Corresponding author: Tel. (630) 252-5471; Fax. (630) 252-7777
E-mail address: hinks@anl.gov

INTRODUCTION

Chemical control of the superconducting properties of $MgB_2$ has proven to be difficult. Significant substitution on either the Mg or B sites has been achieved for only a few elements -- for example, Al on the Mg site [1] or C on the B site [2 - 6]. Even for these elements achieving substitution is not simple. Al doping beyond 10% leads to phase separation [1], and phase transitions have been found at the 17 and 75% doping levels [7]. Mg-Al ordering at 50% doping leads to a $MgAlB_2$ compound with a $T_c$ of 12K [7]. Carbon substitution experiments have yielded conflicting results; from reports of no substitution [2], to phase separated samples [3], to limited solubility of about 15% with different $T_c$ behavior with composition. [4 - 6]  The solubility limits for most other elements have not yet been intensively investigated.  The solubility limit (x in $Mg_{1-x}M_xB_2$) has been reported for Li as 0.3 [8] or 0.15 [9]; for Na as 0.2 [9], for Zn as 0.1 [10], for Mn as 0.03 [10] and for Cu as 0 [11].  There is no case where chemical substitution has increased $T_c$ above the value achieved for high-quality samples of pure $MgB_2$ (39 K).

Based on the literature for transition-metal diboride compounds, one might expect that changing the stoichiometry, for example, creating Mg vacancies, might be another way to alter the superconducting properties of $MgB_2$.  In 1970, Cooper et al. [12] reported that they could raise the $T_c$ of $NbB_2$ to 3.87 K by synthesizing boron-rich compositions near $NbB_{2.5}$. They also reported that boron-rich compositions gave the highest $T_c$, above 11 K, in $Zr_{0.13}Mo_{0.87}B_{2+x}$.  This is the highest $T_c$ reported for a transition-metal diboride. However they made no comment about how they thought the nonstoichiometric composition was accommodated.  More recently, the $(Zr,Mo)B_{2+x}$ system has been reinvestigated using neutron powder diffraction.[13]  It was shown that the nonstoichiometry is accommodated by the formation of vacancies on the Zr/Mo site and that vacancy concentrations up to 15% can be achieved.

Claims of vacancy defect formation on the Mg site in $MgB_2$ have been much less clear. Zhao et al. [14] investigated Mg-deficient compositions and reported that lattice parameters and $T_c$ changed with starting composition.  They explained their results as being due to either Mg vacancies or interstitial B atoms.  However, they also observed $MgB_4$ as an impurity phase in their Mg-deficient samples, which argues against the formation of Mg vacancies.  Gibb's phase rule (for a two-component system) is violated if a variable Mg vacancy content in $MgB_2$ and the expected impurity phase both appear under equilibrium conditions at constant T and P.  Serquis et al. [15] analyzed the Mg vacancy content and lattice strain in a series of $MgB_2$ samples using x-ray Rietveld analysis.  They reported Mg vacancy concentrations up to 5% in their samples and concluded that $T_c$ scaled with both the Mg vacancy concentration and strain.  Single crystal x-ray diffraction of samples grown under high-pressure, high-temperature conditions also showed a 4% Mg vacancy concentration.[16]  Margadonna et al. [17]

found large variations in the phase inhomogeneity in samples synthesized under high-pressure Ar. This variation was modeled with coexisting, Mg deficient, $Mg_{1-}B_2$ phases.

A number of studies have invoked the existence of nonstoichiometry in $MgB_2$ as an explanation for differences among samples without providing experimental evidence that nonstoichiometry actually exists. For example, Chen et al. [18] found a correlation between the residual resistance ratio and magnetoresistance for sample made at different starting compositions, from which they concluded that defect scattering, possibly disorder on the Mg site, was responsible for the sample to sample variation. Bordet et al. [19] suggested that variations in Mg vacancy concentration could explain large differences reported for the pressure dependence of $T_c$. A number of authors have speculated that Mg deficiency could explain the depressed $T_c$'s observed in thin films.[20] Conversely, other authors have concluded that accidental impurity doping was a more likely explanation for differences among samples. Recently, Ribeiro et al. [21] investigated the effect of B purity on transport properties. They concluded that impurities in the B incorporated into the $MgB_2$ during synthesis can have a major effect on $T_c$ and on the residual resistivity ratio. They argue that impurities accidentally incorporated during synthesis, not nonstoichiometry, probably account for the differences in sample properties.

The Mg-B phase diagram has not been reliably determined, however the various Mg-B phases are fairly well documented. A representative phase diagram based on the available experimental information is given by reference 22. Liu et al. [23] have calculated the phase diagram for the system and both the experimental and calculated diagrams agree. $MgB_2$ is a line compound in equilibrium with Mg on the Mg rich side of the phase diagram and in equilibrium with $MgB_4$ on the B rich side.

In this paper we investigate the stoichiometry of $MgB_2$ by studying a range of compositions, $Mg_xB_2$, made at the same synthesis temperature. We determine both the phase composition and stoichiometry using neutron powder diffraction and the superconducting transition temperature with ac susceptibility. No variations were observed on traversing the phase boundary between B rich and Mg rich compositions that could be related to any change in stoichiometry of the $MgB_2$ phase. Although small changes in both lattice constants and $T_c$'s were observed with composition, these changes result from strain and impurity effects, not from a variation in stoichiometry. Our conclusion is that at 850°C, $MgB_2$ is a line compound that may be slightly nonstoichiometric with about 1 % Mg vacancies.

SYNTHESIS AND NEUTRON POWDER DIFFRACTION

A series of $Mg_xB_2$ samples was synthesized with starting compositions from x = 0.6 to 1.3. The same B lot (Eagle-Picher $^{11}$B enriched to 99.52%) and high-purity 1/8" diameter Mg rod were used for the synthesis of all samples to avoid any differences in starting impurity contents. The reaction crucible and lid were machined from HBC grade

BN from Advanced Ceramics Corporation. The same 1.5" diameter by 1.75" height BN crucible was used for the synthesis of all the samples. The B powder was placed in the bottom of crucible with short lengths of the Mg rod resting on the B powder. The Mg rod was etched in 10% acetic acid to remove surface oxidization layer, then weighed and placed in the furnace as rapidly as possible. The sample size was kept at 5g so that any temperature gradients in the sample during synthesis would be the same. The samples were heated to 850°C in 2 hr, held at 850°C for 2 hr followed by furnace cooling. During synthesis, the furnace contained 99.99%Ar at 50 bar pressure.

Neutron powder diffraction data were collected at room temperature on the Special Environment Powder Diffractometer (SEPD) [24] at Argonne's Intense Pulsed Neutron Source. Each diffraction sample consisted of the full contents of the synthesis run, with special care to include any Mg metal that may have condensed at the bottom of the BN synthesis crucible. Data were collected for about 2 hours per sample, with the samples contained in vanadium cans. Care was taken to maintain identical sample mounting for each run in order to minimize systematic errors in the determination of lattice parameters.

Structural refinements were performed using the GSAS code.[25] All refinements were done with the same refinement model, which included contributions from $MgB_2$, Mg, and $MgB_4$. The refined scale factors were used to calculate the weight fractions of the three phases in each sample.

The $MgB_2$ phase was fit using the hexagonal space group P6/mmm.[26,27] Refined parameters included the lattice parameters, the occupancy of the Mg site, and anisotropic temperature factors for both Mg and B. Peak broadening of the $MgB_2$ peaks, up to about twice the instrumental resolution, was observed in the early refinements for some samples. It was determined that this broadening could best be modeled using a Gaussian isotropic strain broadening term, sig-1 in GSAS, plus a Lorentzian anisotropic strain broadening term, g1ec in GSAS, to model a small additional strain parallel to the c axis.

The Mg phase was modeled using space group P63/mmc with Mg at the special position (1/3,2/3,1/4). For samples with a significant contribution from Mg, the lattice parameters and isotropic temperature factor of Mg were refined. For samples with smaller (or no) contribution from Mg, only the scale factor was refined. No significant peak broadening was observed for Mg.

The $MgB_4$ phase was modeled using orthorhombic space group Pnma, with Mg at starting position (0.053,3/4,0.136) and three B atoms at starting positions (0.277,1/4,0.345), (0.442,1/4,0.146),and (0.133,0.559,0.436).[28] In samples with a significant fraction of $MgB_4$, it was possible to refine lattice parameters, atom positions, temperature factors, and an isotropic strain broadening parameter for this phase. For samples with smaller (or no) $MgB_4$ contribution, structural parameters were held fixed in order to achieve stable refinements and only the scale factor of this phase was refined.

RESULTS AND DISCUSSION

An initial series of four samples synthesized using the same heating rate but different reaction times at 850°C was studied to investigate the kinetics of the reaction. The reaction times were 0,1,2 and 8hr. A 2% excess of Mg was used in the synthesis. Figure 1 shows the refined amount of $MgB_4$ in the four samples. The amount of $MgB_4$ reaches a minimum at about 1 to 2 hours, than increases again. The amount never reaches zero as would be expected. Unreacted Mg was observed at the bottom of the crucible for all of the synthesized samples. Mg was also observed in the diffraction pattern of the synthesized samples for the 2 and 8hr reaction times showing that at longer times Mg was also present in the product as well as at the bottom of the crucible. The thermal gradients in the furnace are responsible for transporting the Mg from the initial position on the surface of the B powder to the coldest point at the bottom of the crucible. Some fraction of the Mg has probably transported to the bottom of the crucible before the temperature has increased to a point where reaction kinetics become rapid. Once the final temperature is reached, the reaction mixture is depleted in Mg since it is reacting faster than it can diffuse from the bottom of the crucible. At longer times Mg is again observed in the reaction mixture since the equilibrium vapor pressure of Mg is reached in the product.

The reason the $MgB_4$ phase fraction never goes to zero could possibly be due to either 1), the slight loss of Mg due to volatility or its reaction with residual oxygen in the furnace or 2), the decomposition of $MgB_2$ due to the thermal gradients in the sample. The decomposition pressures for $MgB_2$ have been calculated by Liu et al.[23] Using their result, a thermal gradient of about 400°C would be required to have $MgB_4$ at the hotter sample top and Mg at the cooler crucible bottom. This thermal gradient is not possible, thus the increasing $MgB_4$ content is most likely due to the lower Mg vapor pressure at the top surface of the sample due to reaction with residual oxygen in the furnace or Mg volatility. This conclusion is supported by the observation of black $MgB_4$ present on the top surface of the reaction product after the 8hr reaction time.

The overall composition of the sample is thus determined by two rate constants. There is a rapid synthesis step as the Mg reacts with the B that, from Fig. 2, would likely be completed in about 2hr under the conditions used for this work. There is also a slow decomposition step leading to $MgB_4$ that is determined by the rate of Mg loss from the top surface of the reaction mixture. This leads to a slightly nonuniform sample with Mg at the bottom of the crucible and $MgB_4$ at the top surface of the sample.

The $T_c$ 's (measured at the 5% level of the diamagnetic ac susceptibility) of the 0,1,2 and 4 hr samples were 39.21, 39.15, 39.23 and 39.08(5)K, respectively. The $T_c$ was only significantly reduced for the sample with the longest reaction time showing the most $MgB_4$. This consistency of the $T_c$ indicates the same stoichiometry and purity level of the $MgB_2$ in the reaction product.

Although these results show that the synthesis method is not perfect, they confirm that problems with Mg loss through volatility and secondary reactions with furnace impurities are sufficiently well controlled to allow the synthesis of high quality samples over a range of reaction times; i.e., the synthesis method can be viewed as leading to a final product in equilibrium conditions. This allows phase rules to be invoked in interpreting the data. A reaction time of two hours was used for the composition dependant studies. This is a compromise between a time long enough to completely react the sample and short enough to minimize the amount of $MgB_4$.

Figure 2 shows the refined amount of impurity (Mg or $MgB_4$) present in a series of $MgB_2$ samples as a function of Mg starting composition after a 2hr reaction time. No MgO was found in the diffraction analysis. This places an upper limit of about 2% on the amount of any oxygen contamination of the sample. Any Mg present at the bottom of the crucible was also transferred into the sample container for diffraction analysis. As expected, the dominate impurity in the Mg rich samples is Mg and in the Mg deficient samples is $MgB_4$. However, for the reasons discussed above, there is no point around $x=1$ at which a single phase is found. The extrapolation of the Mg and $MgB_4$ concentrations to zero should show the single phase composition or single phase region of the phase diagram. The least squares fits of the impurity contents extrapolated to zero concentration give $Mg_{1.00}B_2$ and $Mg_{1.01}B_2$ for Mg deficient and Mg excess starting compositions, respectively. Based on only these data, the single phase composition could be on the slightly Mg rich side of the phase diagram or, conversely, a small region of variable nonstoichiometry could exist with up to 1% excess of Mg (or a deficiency of B). However, the errors in this analysis are on the order of a few %, thus, $MgB_2$ appears to be a stoichiometric compound from this chemical analysis of the reaction products.

Rietveld analysis of the site occupancy in these samples indicates about a 1% deficiency in Mg (or excess B) as shown in Fig. 3. The site occupancy is constant, within our experimental accuracy, over the entire phase range investigated. The refined value for the Mg site occupancy is not conclusive evidence that Mg vacancies actually exist. In the Rietveld refinement, Mg site deficiency could be mimicked by chemical substitution on either the Mg or B site (by elements with appropriately different scattering lengths) or by a failure of the Rietveld refinement code to correctly model the anharmonic thermal motion of the B atoms. The latter problem would be expected to yield artificially increased values for both the site occupancy and harmonic temperature factor of B, consistent with our observation of apparent Mg deficiency. Such systematic errors could explain the disagreement between the refined Mg-site occupancy (Fig. 3) and the slight Mg excess found in the chemical phase analysis for the single phase, or single phase region, of the phase diagram (Fig. 2). However, the Mg excess required for the single phase composition found in the phase analysis can also be accounted for by a slight loss of Mg during the synthesis. This could occur due to volatilization and (or) reaction with oxygen (as was shown to occur in the time dependant investigation). Our overall conclusion is that a vacancy content of up to 1% at the Mg site cannot be ruled out. More importantly, the refined Mg site occupancy does not vary with starting composition; in particular there is no rapid change in site occupancy at or near $x=1$ that would indicate solid solution behavior over a small range of starting compositions. Thus, at this temperature, $MgB_2$ is a line compound.

Figure 4a shows the lattice constants as a function of x for the various samples and Fig. 4b shows the c/a ratio. The large scatter in the data points for the a-and c-axis lattice constants, much larger than the statistical errors of the profile fitting as measured by the size of the lattice points, is due to random instrumental errors, e.g. sample positioning. These errors are removed for lattice constant ratios, thus the c/a ratio show in Fig. 4b has higher accuracy than the individual lattice constants. There is a rapid drop in c and c/a across the single phase region around x=1. The a-axis shows an almost linear dependence on Mg content. The c-axis changes by almost a factor of 3.5 more than the a-axis, thus the c/a ratio mimics the change in the c-axis. This rapid decrease in c/a at the nominal single phase composition could be interpreted as evidence for a Mg-vacancy solid solution region in the material, if one ignores the constant refined value for the Mg site occupancy. However, as will be shown, this anomaly in lattice constants is due to lattice strain, not any intrinsic stoichiometry variation in the $MgB_2$.

The profile refinement of the diffraction peak shape for the samples did not show any evidence for particle size broadening, however, strain broadening was observed. The diffraction peak shapes were modeled in terms of an isotropic Gaussian strain broadening term (sig-1) plus an anisotropic strain broadening term (g1ec) that defines a small additional Lorentzian broadening along the c axis (These terms are defined in the GSAS manual.[24]). Figure 5 shows both the isotropic stain parameter, sig-1, and the anisotropic strain parameter, g1ec, for the samples. The c/a ratio and the c-axis lattice constant scale with the much larger isotropic strain broadening (sig-1). The behavior of the lattice parameters and peak broadening versus starting composition can be readily understood by realizing that the behavior divides into three smoothly connected regimes; a Mg rich region, a nearly stoichiometric region and a $MgB_4$ rich region.

In the Mg rich region, above $Mg_{1.1}B_2$, the strain, both isotropic and anisotropic, is lowest and the lattice parameters display their intrinsic values. In this region, where a significant amount of Mg metal is present as an impurity phase, grain-interaction stresses among the randomly oriented $MgB_2$ crystallites are relieved. Our hypothesis is that some of the Mg metal is distributed on the grain boundaries that separate individual crystallites, allowing stress relaxation on these boundaries. We note a report of Mg in grain boundaries of sintered $MgB_2$ samples observed by analytical electron microscopy, consistent with this hypothesis.[29]

In the intermediate regime, from $Mg_{0.9}B_2$ to $Mg_{1.1}B_2$, the amount of impurity phase, whether Mg metal or $MgB_4$, is small enough that the behavior is dominated by grain interaction stresses among the randomly oriented $MgB_2$ crystallites. Because of these stresses, upon cooling from the synthesis temperature, the $MgB_2$ crystallites are inhibited from realizing the full anisotropy of the intrinsic thermal expansion, which is twice as large along the c axis than along the a axis [30]. Instead, the c-axis thermal expansion is slightly reduced and the a-axis thermal expansion is slightly increased. This yields a slightly larger c axis and slightly smaller a axis, when compared to the strain-free values observed in the strain-free Mg-rich regime.

In the composition region below $Mg_{0.9}B_2$, where the concentration of $MgB_4$ as an impurity phase is high, the material displays the behavior of an $MgB_2/MgB_4$ composite in which the response of $MgB_2$ lattice and peak broadening parameters depends on the thermal expansion behavior of both components of the composite.

We, thus, conclude that the observed small variations in lattice parameters over the range of starting compositions studied results from the multiphase behavior of the samples and the resulting effects on grain interaction stresses, not on any variation of the stoichiometry of the $MgB_2$ phase.

The $T_c$'s for the samples are shown in Fig. 6a. The transition temperatures remain relatively constant in the Mg deficient region of the phase diagram but decrease systematically in the Mg rich region. The overall change in $T_c$ is only 0.3K and there is no rapid change at x =1 that might indicate a solid solution region for Mg vacancies over a narrow range of starting compositions. In Fig. 6b the transition temperatures are plotted versus the calculated phase fraction of $MgB_2$ based on a stoichiometric x=1 phase composition in order to more clearly show the difference between the Mg deficient and Mg rich regions. For the Mg-deficient samples with $MgB_4$ as the second phase, $T_c$ is essentially constant. The line through these data in Fig. 6b is a least squares fit which is nearly horizontal indicating no variation of $T_c$ with phase fraction. In the Mg-rich region of the phase diagram, $T_c$ decreases continuously with decreasing $MgB_2$ phase fraction. This behavior can't be explained by any variation in the intrinsic nonstoichiometry of the $MgB_2$ phase. Additionally, there is no correlation between lattice strain and $T_c$; indeed, the Mg-rich region that shows a small, constant lattice strain has the largest change in $T_c$. One must conclude that $T_c$ is independent of strain.

A simple explanation for the observed $T_c$ dependence of the Mg content of the sample is that an impurity (or impurities) in the Mg incorporate into the $MgB_2$ phase and lower its transition temperature. Figure 7 shows a hypothetical phase diagram that could account for the observed $T_c$ behavior. In this figure SS indicates the solid solution region for the impurity (Im) in a given phase and SL is the impurity solubility limit for that phase. The heavily lined region on the Mg-B composition line is the composition region investigated in this study. The two solid circles on the $Mg_{1-\delta}Im_\delta$ - B composition line show the region in phase space traversed in this study assuming some arbitrary impurity Im with a content $\delta$ in the starting Mg.

The behavior divides into three regions. In the Mg + $MgB_2$ (SS) phase region

$$xMg_{1-\delta}Im_\delta + 2B \rightleftharpoons Mg_{1-x}Im_x B_2 + (x-1)Mg \qquad (1)$$

for x ≈ 1 and assuming that the SS region for $MgB_2$ extends to x. In this region, the impurity content of the $MgB_2$ will increase linearly with Mg content and, assuming a linear dependence of $T_c$ with x, will lead to a linear suppression of $T_c$ as is observed. Note that if the impurity responsible for the $T_c$ suppression was coming from the B, $T_c$ would remain constant in this region.

In the $MgB_4(SL) + MgB_2(SS)$ region the impurity content will decrease to zero as the Mg content decreases. Quantitatively,

$$x\, Mg_{1-\alpha}Im_\alpha + 2B \rightleftharpoons (2x-1)Mg_{1-\alpha'}Im_{\alpha'}B_2 + (1-x)Mg_{1-c}Im_cB_4 \qquad (2)$$

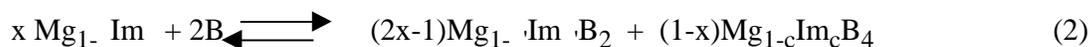

where c is the solubility limit for I in $MgB_4$ and $\alpha' = (c(1-x) - x\alpha)/(1-2x)$. Equation (2) is valid for x = 1 to x = c/(α + c), the starting composition where the impurity content of the $MgB_2$ phase goes to zero and the $T_c$ becomes maximum. Thus the $T_c$ maximum should occur in this two-phase, Mg deficient region. Although the errors in the $T_c$ determination are large, the data in Fig. 6a might suggest that the maximum $T_c$ does indeed occur at x < 1.

In the $MgB_2 + MgB_4(SS)$ region of the phase diagram $T_c$ should remain constant since the impurity content would remain at zero and all of the impurity would be incorporated into the $MgB_4$ phase. This is consistent with the data in Fig. 6b, which show a constant $T_c$ within the accuracy of the measurement. The $T_c$ behavior would be quite different if B had been the host for the impurity. $T_c$ would be high and constant in the $MgB_2 + MgB_4(SS)$ phase region and show a decrease in the $MgB_2(SS) + MgB_4(SL)$ phase region analogous to the case for Mg hosting the impurity. However, in the $Mg + MgB_2(SS)$ phase region $T_c$ would remain constant and suppressed. Thus, in our case the dominant effect on $T_c$ is due to impurity contamination from the starting Mg.

The actual behavior of $T_c$ will depend on the impurities present, their distribution in the starting materials, and the nature of the Mg-B-Im phase diagram for each particular impurity. This could lead to very complex behavior of $T_c$ with starting composition depending on the number of impurities, their effect on $T_c$ and their interaction with the relevant boride phases. In our case, the impurity content was low, and we could model the $T_c$ in a very simple way.

Ribeiro et. al. [21] have shown that the $T_c$ of $MgB_2$ varies in a somewhat consistent manner with the claimed purity of the starting B, with the highest $T_c$ shown for material synthesized from the second - highest - clamed purity B (i.e. isotopic $^{11}B$ from Eagle-Picher, the same as used in this study). Although the Eagle-Picher $^{11}B$ may have more total impurities than the clamed highest purity product, it probably has the least impurities that affect $T_c$. We have done chemical analysis in order to see if any impurities follow the phase behavior that we have suggested could model the $T_c$ behavior. Table 1 shows the inductively coupled plasma-atomic emission spectroscopic (ICP-AES) analysis of the starting materials and 3 samples, two synthesized at the composition extremes and one near the stoichiometric composition. The samples were analyzed at two different times as two separate batches, marked I and II in the table. Although the overall accuracy is only ± 20%, within a given batch the precision is ± 5%. The sample for analysis was taken from the synthesized powder and did not include any of the excess Mg at the bottom of the crucible for the $Mg_{1.3}B_2$ sample. The numbers in parenthesis are calculated based on the analyzed starting compositions and do not include any correction for the Mg phase segregation for the $Mg_{1.3}B_2$ composition. The impurity calculations were done if both starting elements had analyzed impurity concentrations or if the impurity was dominant in only one starting element and we could assume zero concentration in the second element.

The impurity elements group into 3 classes. The first class (Si, Fe) have large, random variations from the expected values. These elements most likely have large nonuniform distributions in the solid phase due to small diffusion constants at the low synthesis temperature. The second group of elements (Ca, Mn, In) have values larger than expected. These elements have most likely been additionally doped into the material from external sources during the synthesis and handling procedures. The last group (Zn, Pb, Al) on the other hand, have concentrations that scale in a manner consistent with the $T_c$ behavior, i.e. increasing impurity content as the Mg starting composition increases. Zn and Pb actually follow the calculated impurity content fairly closely. These elements, being volatile, attain a uniform distribution in the solid phase, however, may not actually have been incorporated into lattice. None of these elements is likely responsible for the $T_c$ variation in the samples, the concentration is far too low. Al, for example, would require almost 6000 ug/g for a 0.2K drop in $T_c$ based on the work reported in Ref. 1. Even though it may be impossible to identify the impurity responsible for the $T_c$ variation, the analysis shows that some dopants do follow the correct phase behavior.

CONCLUSIONS

In conclusion, we have shown that, for the synthesis conditions employed in this work (850°C in 50 atm. Ar gas), $MgB_2$ is a line compound; i.e., it does not have variable composition. The composition may be slightly (up to 1%) Mg deficient; however, other effects such as a failure of the Rietveld refinement to adequately model the anharmonic motion of B or chemical substitution on either the Mg or B site can mimic Mg vacancies in structure refinements. In this regard, we note that chemical substitution may provide an alternative explanation for the 4% Mg vacancy concentration observed in single-crystal x-ray diffraction studies of $MgB_2$ crystals grown at high pressure from the Mg-B-N system.[15] In particular, incorporation of N at the B site would raise the x-ray scattering cross section of that site, leading to an indication of Mg deficiency in a refinement that assumed only B on the B site. A more precise determination of the composition may be possible by exploiting the different scattering contrasts in a joint neutron and x-ray diffraction study or by pyctnometric measurements of the density of single crystals. However, achieving better accuracy than the present study will be difficult.

Although these results apply only to $MgB_2$ made under these synthesis conditions, they are likely to hold for other synthesis methods near the same range of temperature and pressure. Moreover, this study illustrates the approach that must be taken to explore whether variable composition can be achieved by other synthesis methods. First, the synthesis technique must adequately approximate equilibrium conditions. Second, all products of the synthesis must be accounted for in the analysis. Phase rules can only be applied if these two conditions are satisfied.

We have also shown that small variation of lattice parameters with starting composition is not an adequate proof of variable composition in the $MgB_2$ phase. In the present study, small variations arise from grain-interaction stresses, which vary with the phase composition of the sample, leading to a rather complex behavior of the lattice parameters with starting composition. We note that stress effects could be greatly amplified in samples made at high pressure because stresses introduced under high-pressure synthesis conditions, where strong sintering between grains occurs, may not be able to relax.

A small variation of $T_c$ with starting composition observed in the Mg-rich regime in our study can be explained in terms of accidental doping by an impurity in the starting Mg. Accidental impurity doping has been observed previously in $MgB_2$. For example, Ribeiro et al. reported rather large changes in $T_c$ depending on the purity of the starting B.[21] However, in both their study and ours, it has not been possible to identify the impurity or to quantify its concentration in the $MgB_2$. A detailed interpretation of accidental impurity doping effects requires a knowledge of the ternary phase diagram for Mg, B, and that particular impurity. We have shown in the present case that it is possible to hypothesize a phase diagram that is consistent with all experimental observations.

It may be possible to extend the concept of accidental impurity doping to develop methods for achieving deliberate chemical substitution in $MgB_2$. For example, alloys of Mg might be used to incorporate other metals on the Mg site. It has already been shown that BC can be used as a starting material to synthesize C-doped $MgB_{2-x}C_x$.[4] Creative use of such ideas may provide a route for exploring chemically substituted $MgB_2$ in cases where other synthesis routes have failed.

ACKNOWLEDGEMENT

This work was supported by the U.S. Department of Energy, Office of Science, Contract No. W-31-109-ENG-38.

Figure Captions

Figure 1. Weight fraction of $MgB_4$ in $Mg_{1.01}B_2$ samples synthesized at 850°C for different firing times.

Figure 2. Weight fraction of $MgB_4$ and Mg in synthesized samples of $MgB_2$ with different starting Mg/B rations. The lines through the data points are least squares fit to the data. The errors in the weight fractions are smaller than the data symbols.

Figure 3. Refined site occupancy for Mg as a function of the starting Mg/B ratios for synthesized $MgB_2$ samples.

Figure 4. Lattice constants a and c (4a) and the c/a ration (4b) as a function of the Mg/B starting composition for synthesized $MgB_2$ samples. The lines are only a guide for the eye. The statistical error from the profile analysis is smaller than the data points for the a- and c-axis.

Figure 5. The isotropic (sig-1) and anisotropic (g1ec) strain broadening parameters determined from the Rietvelt profile analysis of the neutron powder diffraction pattern for synthesized samples of $MgB_2$ as a function of starting Mg/B composition. The line is only a guide for the eye. The statistical error from the profile analysis is smaller than the data points for the strain broadening parameters.

Figure 6. The superconducting transition temperatures ($T_c$) for different samples of $MgB_2$ synthesized with different starting compositions. Fig. 6a shows $T_c$ vs. x and Fig. 6b shows $T_c$ vs. the calculated phase fraction of either $MgB_4$ (Mg deficient) or Mg (Mg-rich). In Fig 6b the lines are the least-squares fit to the data.

Figure 7. A simple phase diagram for an impurity Im in $MgB_2$. The heavy line on the Mg-B composition line is the composition region investigated in this study. The circles on the $Mg_{1-}$ Im -B composition line delineate the line in phase space containing our samples assuming an impurity content of     for the starting Mg.

Table 1. ICP-AES analysis for the starting B and Mg materials and 3 $Mg_xB_2$ synthesized compositions. Impurity concentrations are reported in µg/g and the numbers in parentheses are calculated values, where possible, based on the analyzed starting values. The excess Mg at the bottom of the synthesis crucible for the $Mg_{1.3}B_2$ sample was not included in the analyzed sample. The samples were analyzed as two different groups at different times marked I and II. The accuracy of the analysis is ± 20%, however, the precision within a given group is ± 5%.

| Sample | Mg (I) | B (II) | $Mg_{0.6}B_2$ (I) | $Mg_{0.95}B_2$ (II) | $Mg_{1.3}B_2$ (I) |
|---|---|---|---|---|---|
| Si | 650 | 320 | 530 (450) | 240 (490) | 330 (513) |
| Fe | 25 | 9 | 28 (15) | 15 (17) | 23 (18) |
| | | | | | |
| Ca | 28 | 26 | 69 (27) | 46 (27) | 42 (27) |
| In | 170 | < 25 | 204 (67) | 140 (87) | 189 (100) |
| Mn | 17 | < 10 | 41 | 15 | 23 |
| | | | | | |
| Zn | 42 | < 10 | < 20 (17) | 15 (21) | 24 (25) |
| Pb | 89 | < 20 | 36 (35) | 46 (45) | 50 (52) |
| Al | 19 | < 40 | 28 | < 40 | 37 |

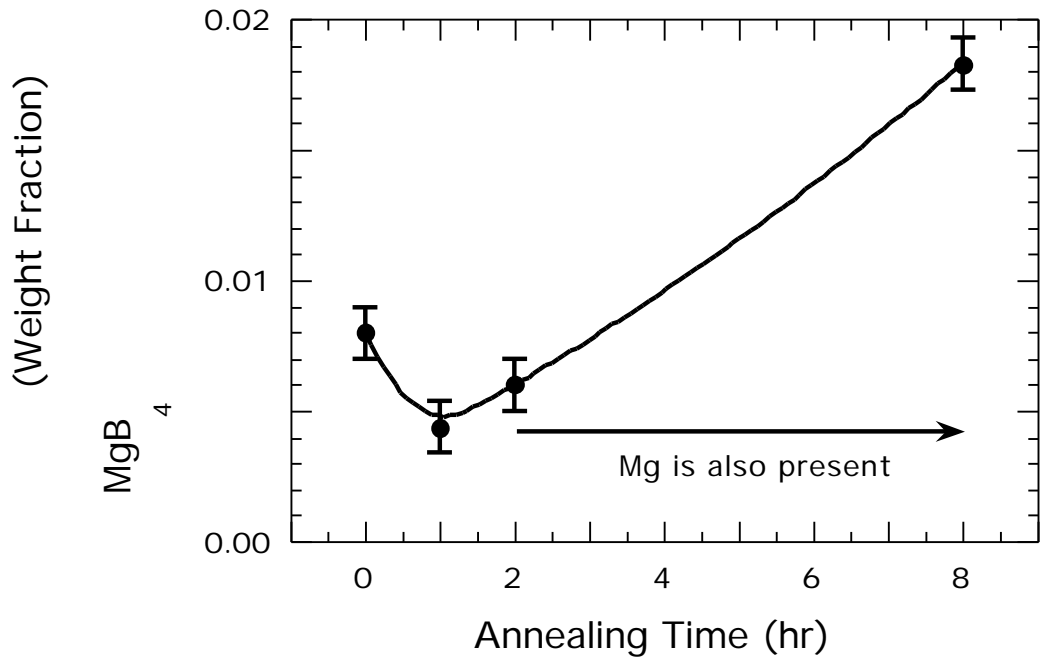

Figure 1

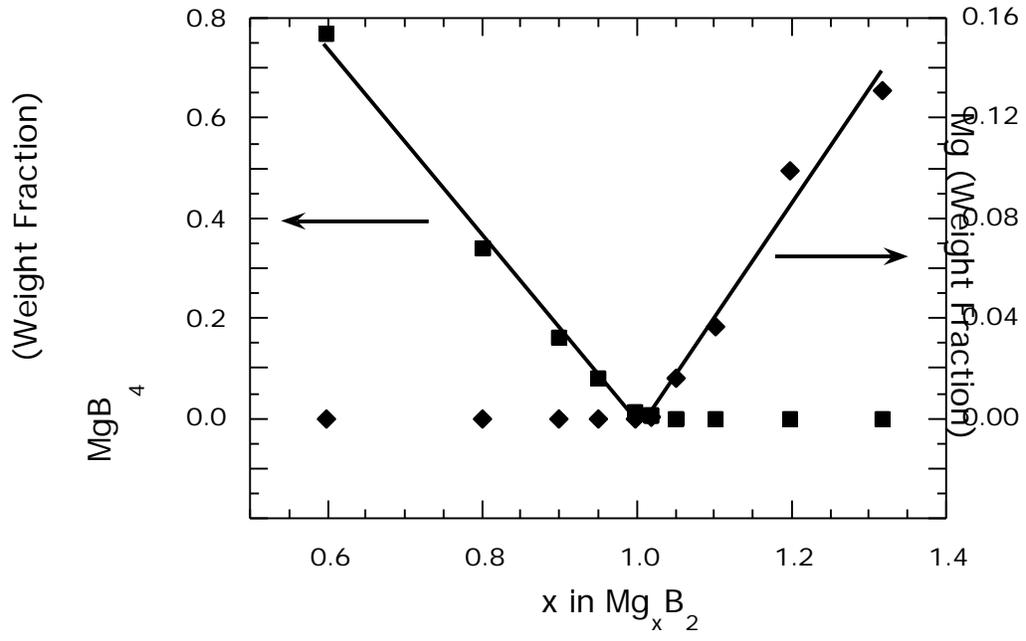

Figure 2

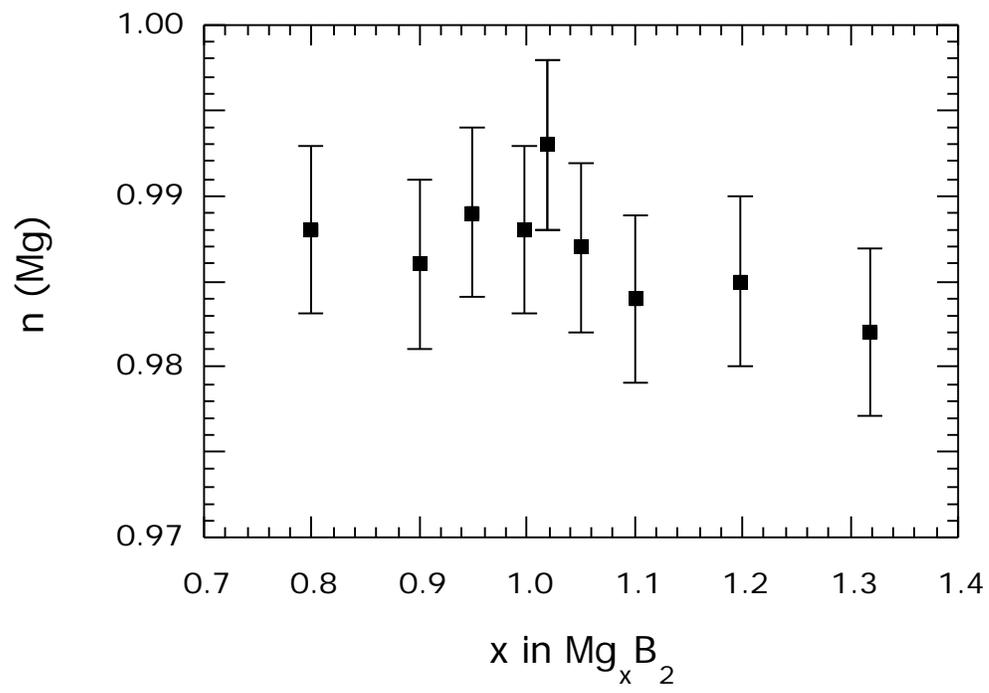

Figure 3

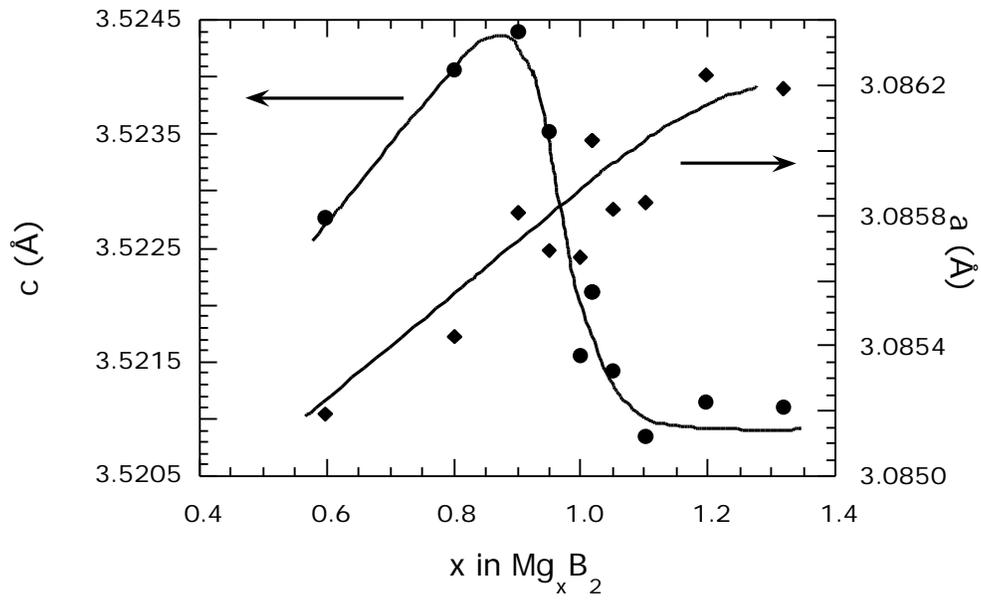

Figure 4a

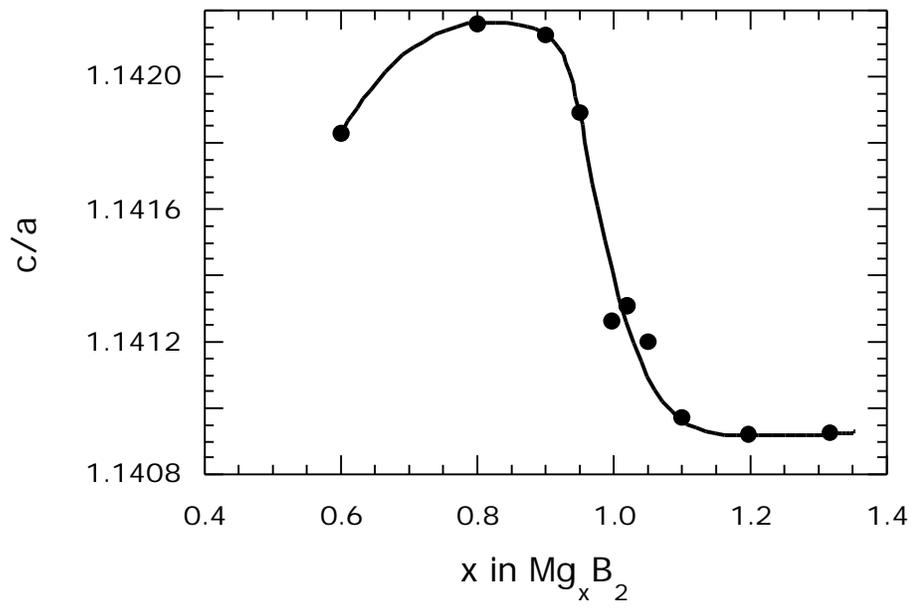

Figure 4b

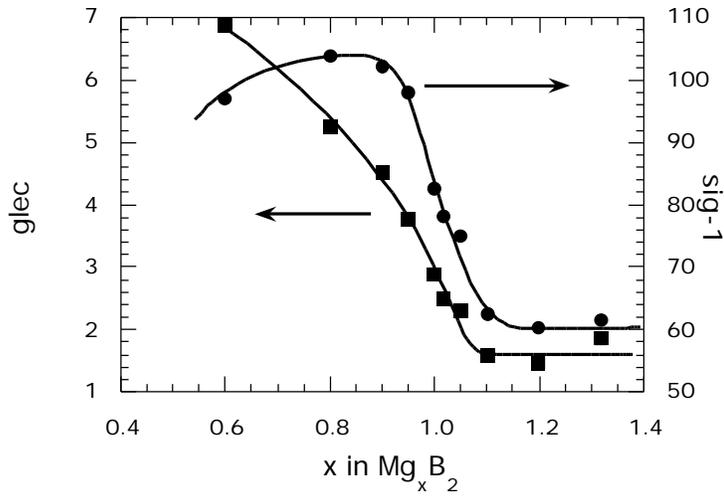

Figure 5

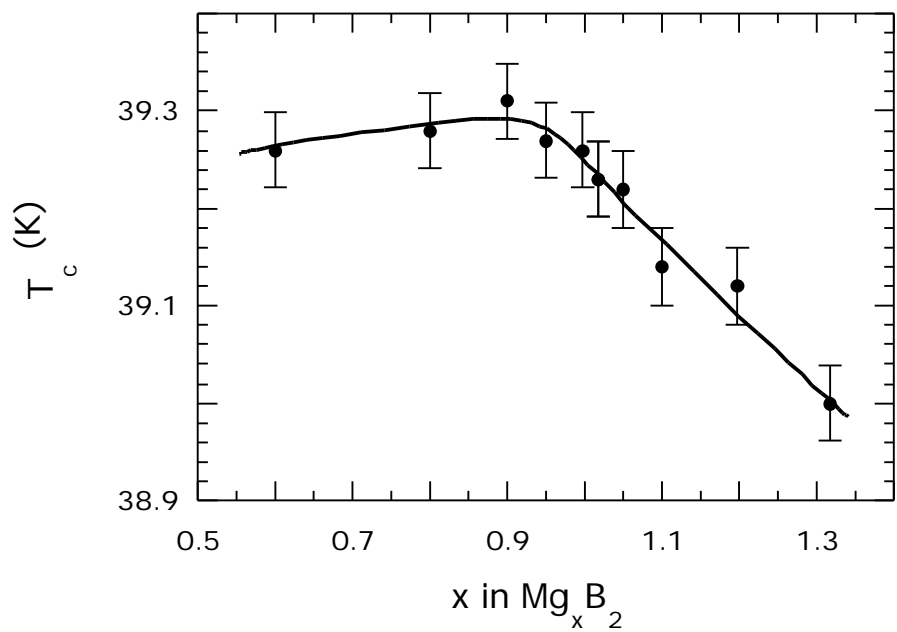

Figure 6a

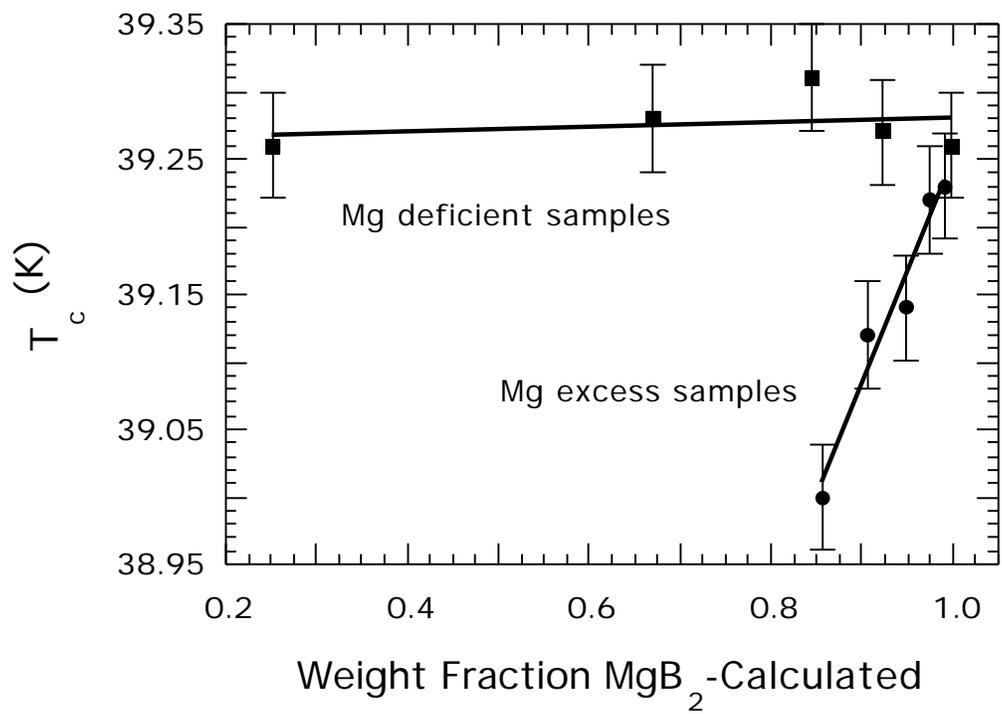

Figure 6b

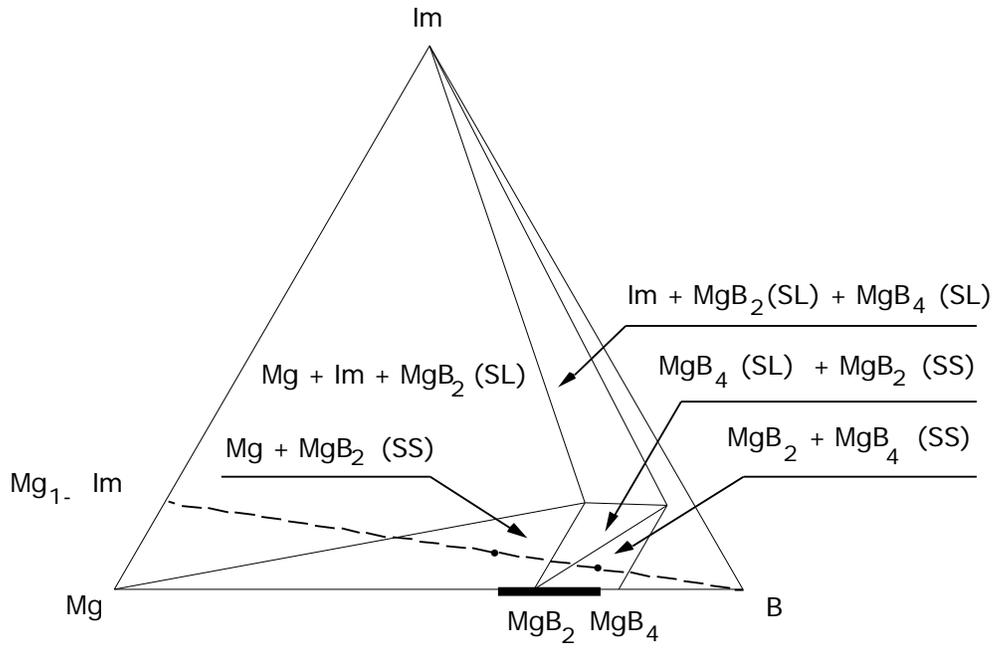

Figure 7